\documentstyle[amssymb,pra,aps,multicol,psfig]{revtex}

\def\TwoColumns{y}

\def\YES{y} 

\if\TwoColumns\YES
  \def\narrowtextAAA{\begin{multicols}{2}}
  \def\widetextAAA{\end{multicols}  }
  \def\widetextEQ{\end{multicols}

     \noindent\vrule width 0.49\textwidth height 0.5pt\vrule width 0.5pt height 5pt\hfill

  }
  \def\narrowtextEQ{

    \noindent\hfill\vrule width 0.5pt height 0.5pt depth 4pt\vrule width 0.49\textwidth height 0.5pt depth 0pt

    \begin{multicols}{2}
  }
  \def\AppendixBEG{
     \begin{appendix}
  }
  \def\AppendixEND{\end{appendix}}
\else
  \def\narrowtextAAA{\narrowtext}
  \def\widetextAAA{\widetext}
  \def\narrowtextEQ{\narrowtext}
  \def\widetextEQ{\widetext}
  \def\AppendixBEG{\begin{appendix}}
  \def\AppendixEND{\end{appendix}}
\fi

\def\mathbf#1{\mbox{\boldmath $#1$}}

\def\vphantomA{\vphantom{ a^{(2)} }}

\begin{document}
\title{Nonclassical correlations in damped quantum solitons}
\author{Eduard~Schmidt, Ludwig~Kn\"{o}ll, and Dirk--Gunnar~Welsch}
\address{Friedrich-Schiller-Universit\"{a}t Jena, 
Theoretisch-Physikalisches Institut\\
Max-Wien Platz 1, D-07743 Jena, Germany}
\date{April 6, 1999}
\maketitle

\begin{abstract}
Using cumulant expansion in Gaussian approximation, 
the internal quantum statistics of damped soliton-like pulses 
in Kerr media are studied numerically, considering
both narrow and finite bandwidth spectral pulse
components. It is shown that the sub-Poissonian statistics 
can be enhanced, under certain circumstances,  
by absorption, which damps out some destructive interferences.
Further, it is shown that both the photon-number correlation
and the correlation of the photon-number variance between
different pulse components can be highly nonclassical
even for an absorbing fiber.
Optimum frequency windows are determined
in order to realize strong nonclassical behavior,
which offers novel possibilities of using
solitons in optical fibers as a source of nonclassically 
correlated light beams.
\end{abstract}



\frenchspacing

\narrowtextAAA


\section{Introduction}
\label{Sec.intro}

Nonclassical features of optical pulses
like squeezing, sub-Poissonian statistics and
entanglement have been of increasing interest for
optical communication and measurement techniques
\cite{DrummondPD93,SizmannA97,AndersonME97,AbramI99}. From
classical optics it is well known that in fibers photons can fly
in soliton-like pulses over long distances, provided that
they intense enough and the fiber nonlinearity can
compensate for the dispersion-assisted pulse spreading
\cite{HasegawaA89,AkhmanovSA92}. Yet solitons are not rigid
but very lively nonclassical species, because of the quantum noise
associated with the nonlinear dynamics. Since already
single-mode radiation becomes nonclassical under the
influence of a Kerr nonlinearity \cite{KitagawaM86}, solitons
can be expected to exhibit not only single-mode nonclassical
properties but also nonclassical internal correlations.

The study of quantum solitons has been also motivated by a number
of possible applications \cite{DrummondPD93,SizmannA97,AbramI99}.
Solitons may be used for realizing efficient and reliable
sources of pulses with intensity fluctuations below the shot
noise level. Nondestructive high-precision optical switching devices
and logic gates may be implemented applying the concept of quantum
non-demolition measurement to the collision interaction of solitons
\cite{FribergSR92,SpaelterS97a,CourtyJM98}. Suppression of noise in
soliton communication lines \cite{DrummondPD93,SizmannA97,AbramI99}
and usage of solitons in quantum information processing
\cite{AbramI99} may be other potential applications.

Nonclassical properties of optical solitons have been
detected in a number of experiments. In \cite{RosenbluhM91,BergmanK94}
soliton squeezing is measured using homodyne detection.
Photon-number squeezing of the spectrally filtered solitons
is measured by direct detection in \cite{FribergSR96,SpaelterS97}.
In the experiment in \cite{SchmittS98}
with asymmetrical ($10/90$) fiber-loop interferometer
photon-number squeezing up to $6.0\mbox{dB}$ is achieved.
Recently the experimental studies have been extended to
internal spectral photon-number correlations associated with
narrow bandwidth soliton components \cite{SpaelterS98}.

Disregarding losses, the dynamics of quantum solitons in optical
fibers may be described by the operator-valued
nonlinear Schr\"odinger equation, which can be solved employing the
Bethe ansatz \cite{LaiY89a}. The method was used successfully to calculate
field and intensity correlations in the space-time domain
\cite{YaoD95a,KartnerFX96}. In practice however, the
quantum nonlinear Schr\"odinger equation must be
modified by supplementing it with further terms in order to
include effects such as absorption, third-order dispersion,
and Raman scattering. Various methods of solution have
been developed and successfully applied
\cite{LaiY89,SingerF92,HausHA90,LaiY90,LaiY93,%
DoerrCR94,LaiY95,DrummondPD87,DrummondPD93a,CarterSJ95,%
WernerMJ96,WernerMJ97,WernerMJ97a,WernerMJ97b,SchmidtE98}.

Here we follow \cite{SchmidtE98} and use cumulant expansion
in Gaussian approximation for studying spectral photon-number
squeezing and nonclassical photon-number correlation of different
spectral components of damped solitons.
We consider both narrow-bandwidth spectral pulse components
\cite{SpaelterS98} and pulse components with finite spectral
bandwidth selected by optimally chosen square bandpass
filters \cite{WernerMJ96}. The numerically obtained results
illustrate the possibility to use optical solitons for generation
of squeezed light and nonclassically correlated light beams
also in the presence of absorption.
Moreover, the results reveal that absorption can improve,
under certain circumstances, the nonclassical features.

The article is organized as follows. Section \ref{Sec.model} outlines
the used basic concept. The relations necessary for studying
nonclassical correlations are given in Sec.~\ref{Sec.CS}, and
the numerical results are reported in Sec.~\ref{Sec.Results}.
Finally, a summary and some concluding remarks are given
in Sec.~\ref{Sec.Summary}.


\section{Basic concept}
\label{Sec.model}

Let us first give a brief outline of the used concept for
describing the propagation of a quantized optical pulse through an
absorbing fiber with second order dispersion and Kerr nonlinearity
(for details, see \cite{SchmidtE98} and references therein).
The spatio-temporal pulse evolution in a co-moving reference frame
is described in terms of slowly varying bosonic operators
$\hat{a}(x,t)$ satisfying the commutation relation
\begin{equation}
\left[ \hat{a}(x,t),\hat{a}^{\dagger }(x^{\prime },t)\right] ={\cal A}%
^{-1}\delta (x-x^{\prime }),  \label{eq.aa}
\end{equation}
[${\cal A}$ is the effective cross-section of the fiber].
The undamped motion is governed by the Hamiltonian
\begin{equation}
\hat{H}=\hbar {\cal A}\int dx\left[ \vphantomA
\textstyle\frac{1}{2}\omega ^{(2)}\left( \partial
_{x}\hat{a}^{\dagger }\right) \left( \partial _{x}\hat{a}\right)
+{\textstyle\frac{1}{2}}\chi
\hat{a}^{\dagger }\hat{a}^{\dagger }\hat{a}\hat{a}\vphantomA
\right] ,  \nonumber
\end{equation}
where the constant ${\chi }$ is related to the third-order susceptibility
$\chi^{(3)}$ as
\begin{equation}
{\chi }=\frac{3\chi ^{(3)}\hbar (v_{{\rm gr}} k_{{\rm c}})^{2}}
{4\epsilon _{{\rm r}}^{2}\epsilon _{0}}  \label{eq.chi}
\end{equation}
[$\epsilon _{{\rm r}}$, relative permittivity at the carrier frequency
$\omega _{{\rm c}}$; $k_{{\rm c}}$, carrier wave number; $v_{{\rm gr}}$,
group velocity;  $\omega ^{(2)}$ $\!=$
$\!{\rm d}^2\omega (k)/{\rm d}k^2|_{k=k_{\rm c}}$].
Note that solitons can be formed either
in focusing media with anomalous dispersion
(\mbox{$\chi$ $\!<$ $0$}, \mbox{$\omega^{(2)}$ $\!>$ $\!0$}) or
in defocusing media with normal dispersion
(\mbox{$\chi$ $\!>$ $\!0$}, \mbox{$\omega^{(2)}$ $\!<$ $\!0$}).

The damped motion is treated on the basis of the master equation
\begin{equation}
i\hbar ~\partial _{t}\hat{\rho}=[\hat{H},\hat{\rho}]
+i\gamma \hat{L}\hat{\rho},  \label{eq.master}
\end{equation}
where
\begin{eqnarray}
\gamma \hat{L}\hat{\rho} &=&\gamma \hbar {\cal A}\int dx\left[ \vphantomA N_{%
{\rm th}}\left( 2\hat{a}^{\dagger }\hat{\rho}\hat{a}-\hat{\rho}\hat{a}\hat{a}%
^{\dagger }-\hat{a}\hat{a}^{\dagger }\hat{\rho}\right) \right.
\nonumber\\
&&\quad \quad \quad +\left. \left( N_{{\rm th}}+1\right) \left( 2\hat{a}\hat{%
\rho}\hat{a}^{\dagger }-\hat{\rho}\hat{a}^{\dagger }\hat{a}-\hat{a}^{\dagger
}\hat{a}\hat{\rho}\right) \vphantomA\right],  \label{eq.L}
\end{eqnarray}
with $\gamma $ being the damping constant, and
\begin{equation}
N_{{\rm th}}=\left[ \exp \left( \frac{\omega _{{\rm c}}\hbar }{k_{{\rm B}}T}%
\right) -1\right] ^{-1}  \label{Nthermisch}
\end{equation}
($k_{{\rm B}}$ - Boltzmann constant, $T$ - temperature). The model
applies to pulses longer than $1\,$ps, otherwise the influence of
additional effects such as Raman scattering and third order dispersion
must be taken into account.

The master equation (\ref{eq.master}) is converted,
after spatial discretization, into a pseudo-Fokker-Planck equation
for an $s$-parametrized multi-dimensional phase-space
function, which is solved numerically using
cumulant expansion in Gaussian approximation.
The initial condition is realized by a multimode displaced
thermal state, without internal entanglement,
and it is assumed that the field expectation value
corresponds to the classical fundamental soliton.
In the numerical calculation, the coordinates $x$ and $t$ are
scaled by the initial pulse width $x_{0}$ and the dispersion
time $t_{{\rm d}}=x_{0}^{2}|\omega ^{(2)}|^{-1}$ respectively.
The calculations are performed
on a grid of $200$ points with discretization step 
$\Delta x$ $\!=$ $\!0.1\,x_0$
for an initial photon number of the
pulse of $2\bar{n}$ $\!=$ $\!2|\omega ^{(2)}{\cal A}/(\chi x_{0})|$
$\!=$ $2\!\cdot\!10^{9}$ and a reservoir photon number of $N_{{\rm th}}$ 
$\!=$ $\!10^{-16}$. The value of $N_{{\rm th}}$ corresponds to a
carrier wavelength
of $\lambda _{{\rm c}}$ $\!=$ $\!2$ $\!\!\pi $ $\!\!c$ $%
\!\!/\omega _{{\rm c}}$ $\!=$ $1.5\,\mu {\rm m}$ of the pulse in vacuum and
a temperature of $T$ $\!=$ $\!300$K [see Eq.~(\ref{Nthermisch})]. Thus the
pulse is initially prepared in a displaced thermal state that is almost a
coherent state.


\section{Nonclassical spectral correlations}
\label{Sec.CS}

In order to study spectral properties, bosonic operators in
the Fourier space are introduced,
\begin{equation}
\label{eq.aw}
\hat{a}(\omega ,t)=\frac{1}{\sqrt{2\pi }}\int_{-\infty }^{\infty }dx\
e^{i\omega x}\hat{a}(x,t).
\end{equation}
The operator $\hat{N}_i$ of the number of photons
in a frequency interval
$(\Omega_i,\Omega_i^{\prime })$, i.e.,
$\Omega_i$ $\!\leq$ $\!\omega$ $\!\leq$ $\!\Omega_i^{\prime}$, reads
\begin{equation}
\label{eq.NWW}
\hat{N}_i={\cal A}\int_{\Omega_i }^{\Omega_i ^{\prime
}}d\omega \ \hat{a}^{\dagger }(\omega ,t)\hat{a}(\omega ,t),
\end{equation}
and the photon-number variance of the beam associated with this
frequency interval can be given by
\begin{equation}
\langle \Delta \hat{N}_{i}^{2}\rangle =\langle \hat{N}_{i}^{2}\rangle
-\langle \hat{N}_{i}\rangle ^{2}
=\langle :\!\Delta \hat{N}_{i}^{2}\!:\rangle
+\langle \hat{N}_{i}\rangle ,
\label{eq.dn}
\end{equation}
where $:\ :$ introduces normal ordering.
When the inequality
\begin{equation}
\langle :\!\Delta \hat{N}_{i}^{2}\!:\rangle < 0
\label{eq.dn1}
\end{equation}
is valid -- a condition that cannot be satisfied within the
classical noise theory --  then sub-Poissonian statistics are
observed, i.e., $\langle \Delta \hat{N}_{i}^{2}\rangle$ $\!<$
$\!\langle \hat{N}_{i}\rangle $.

Let us now consider two nonoverlapping frequency intervals
$(\Omega _{i},\Omega _{i}^{\prime })$, $i$ $\!=$ $\!1,2$.
A measure of the mutual correlation of the photon-number variances
in the two frequency intervals is the correlation coefficient
\begin{equation}
\eta_{12}=\frac{\langle \Delta\hat{N}_{1} \Delta\hat{N}_{2}\rangle }
{\sqrt{\langle \Delta\hat{N}_{1}^{2}\rangle
\langle \Delta\hat{N}_{2}^{2}\rangle }} \,.
\label{eq.eta12}
\end{equation}
(note that $\langle \Delta\hat{N}_{1} \Delta\hat{N}_{2}\rangle$
$\!=$ $\!\langle :\!\Delta\hat{N}_{1} \Delta\hat{N}_{2}\!:\rangle$
for nonoverlapping frequency intervals). From the Cauchy-Schwarz inequality
\begin{equation}
\langle \Delta\hat{N}_{1}^{2}\rangle \langle \Delta\hat{N}_{2}^{2}\rangle
- \langle \Delta\hat{N}_{1} \Delta\hat{N}_{2}\rangle^{2} \geq 0
\label{eq.CS.op}
\end{equation}
it follows that the absolute value of the correlation coefficient
is limited to the right,
\begin{equation}
|\eta _{12}|\leq 1.  \label{eq.eta12.1}
\end{equation}
Obviously, the correlation is nonclassical, if
the inequality (\ref{eq.CS.op}) is violated for
normally ordered quantities, i.e.,
\begin{equation}
\langle :\!\Delta \hat{N}_{1}^{2}\!:\rangle
\langle :\!\Delta \hat{N}_{2}^{2}\!:\rangle
-\langle :\!\Delta \hat{N}_{1}\Delta \hat{N}_{2}\!:\rangle^{2} < 0.
\label{eq.CS.dn}
\end{equation}
In order to give a quantitative measure of the degree of
nonclassical correlation, we define the generalized correlation
coefficient
\begin{eqnarray}
\label{eq.y12}
\tilde{\eta}_{12}
&=&\frac{\langle :\!\Delta \hat{N}_{1}^{2}\!:\rangle
\langle :\!\Delta \hat{N}_{2}^{2}\!:\rangle
-\langle :\!\Delta \hat{N}_{1}\Delta \hat{N}_{2}\!:\rangle^{2}}
{\left| \langle \Delta \hat{N}_{1}^{2}\rangle
\langle \Delta \hat{N}_{2}^{2}\rangle -
\langle :\!\Delta \hat{N}_{1}^{2}\!:\rangle
\langle :\!\Delta \hat{N}_{2}^{2}\!:\rangle \right| }
\nonumber\\
&=&\frac{\eta _{11}\eta _{22}-\eta _{12}^{2}}
{|1-\eta _{11}\eta _{22}|}
\end{eqnarray}
where
\begin{equation}
\eta _{ii}=
\frac{\langle :\!\Delta \hat{N}_{i}^{2}\!:\rangle }{\langle \Delta
\hat{N}_{i}^{2}\rangle }\leq 1.  \label{eq.eta11}
\end{equation}
Note that $\eta _{ii}$ is
negative (positive) for sub-Poissonian (super-Poissonian) statistics.
Nonclassical correlation is observed if
\begin{equation}
\label{eq.eta111}
\tilde{\eta}_{12} < 0,
\end{equation}
and it is easily proved that
\begin{equation}
\label{eq.y121}
\tilde{\eta}_{12} \ge -1.
\end{equation}
   From Eq.~(\ref{eq.y12}) it is easily seen that when
\begin{equation}
\langle :\!\Delta \hat{N}_{1}^{2}\!:\rangle
\langle :\!\Delta \hat{N}_{2}^{2}\!:\rangle <0,
\label{eq.y12.1}
\end{equation}
then the correlation of the photon-number variances is
nonclassical. Note that for Gaussian quantum states the
criterion (\ref{eq.CS.dn}) is closely related to that
used in \cite{SpaelterS98}.

Similarly, the mutual correlation of the numbers of
photons in the two frequency intervals can be considered,
\begin{equation}
\tau_{12}=\frac{\langle \hat{N}_{1} \hat{N}_{2}\rangle }
{\sqrt{\langle \hat{N}_{1}^{2}\rangle \langle \hat{N}_{2}^{2}\rangle }}\,
\label{eq.eta121}
\end{equation}
($|\tau_{12}|$ $\!\leq 1$). Nonclassical correlation is realized if
\begin{equation}
\langle :\!\hat{N}_{1}^{2}\!:\rangle \langle :\!\hat{N}_{2}^{2}\!: \rangle
- \langle :\!\hat{N}_{1} \hat{N}_{2}\!: \rangle ^{2} < 0 ,
\label{eq.CS.op1}
\end{equation}
or equivalently
\begin{equation}
\label{eq.CS.op2}
\tilde{\tau}_{12} < 0 ,
\end{equation}
with
\begin{equation}
\tilde{\tau}_{12}
=\frac{\langle :\!\hat{N}_{1}^{2}\!:\rangle
\langle :\!\hat{N}_{2}^{2}:\rangle
-\langle :\!\hat{N}_{1}\hat{N}_{2}\!:\rangle ^{2}}
{\langle \hat{N}_{1}^{2}\rangle
\langle \hat{N}_{2}^{2}\rangle
-\langle :\!\hat{N}_{1}^{2}\!:\rangle
\langle :\!\hat{N}_{2}^{2}\!:\rangle }
\label{eq.Y12}
\end{equation}
($\tilde{\tau}_{12}$ $\!\ge$ $\!-1$).
Note that for the photon numbers an inequality analogous to
(\ref{eq.y12.1}) is not valid, since the average of the
normally ordered square of a photon number cannot be negative.

\begin{figure}[tbp]
\psfig{file=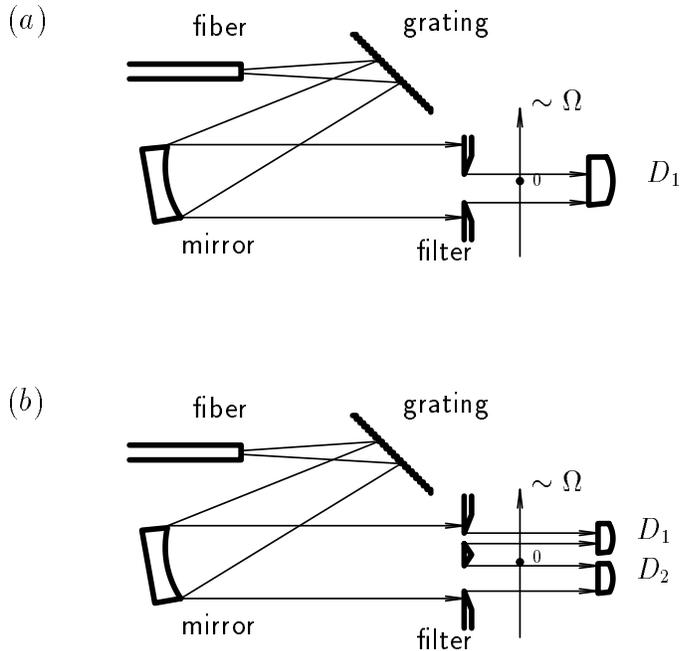,width=10cm,clip=}
\caption{
  Experimental setup for studying the spectrally resolved
quantum statistics of soliton pulses.
The scheme $(a)$ is suited for measurement of the
photon-number statistics \protect\cite{FribergSR96,SpaelterS97},
whereas the scheme $(b)$ allows one to perform
correlation measurements. If the detectors $D_1$ and $D_2$
are omitted, the schemes can be used for selecting from the
original beam partial beams that are highly nonclassical.
}
\label{fig.setup}
\end{figure}

The spectral photon-number statistics can be determined using
a setup shown in Fig.~\ref{fig.setup}.
The scheme in Fig.~\ref{fig.setup}$(a)$ is used in \cite{FribergSR96}
(and with slightly modified detection in \cite{SpaelterS97})
for measuring spectral photon-number squeezing. The
scheme can also be used to reconstruct correlations
between different spectral components from the measured data
\cite{SpaelterS98}. However with regard to direct
correlation measurement, a scheme of the
type shown in Fig.~\ref{fig.setup}$(b)$ may be better suited
for that purpose. Omitting the detectors, the schemes can
also be used for selecting from the original beam partial
beams that are highly nonclassical.


\section{Results}
\label{Sec.Results}
\subsection{Narrow bandwidth intervals}
\label{narrow}

\begin{figure}[tbp]
\psfig{file=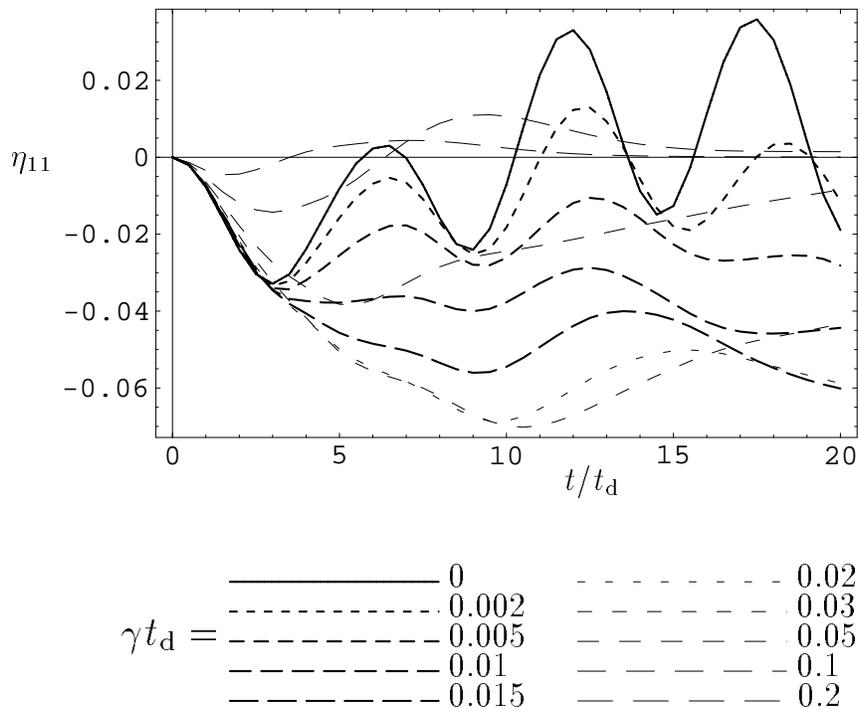,clip=}
\caption{
The temporal evolution of the coefficient $\eta_{11}$,
Eq.~(\protect\ref{eq.eta11}), is shown
for the spectral interval $(\Omega _{1},\Omega _{1}^{\prime })$
$\!=$ $\!(-\Delta\omega,\Delta\omega)$,
\mbox{$\Delta\omega$ $\!=$ $\!0.05\pi\omega_0$}, and
various values of the damping parameter $\gamma$.
}
\label{fig.nn0t}
\end{figure}

\begin{figure}[tbp]
\psfig{file=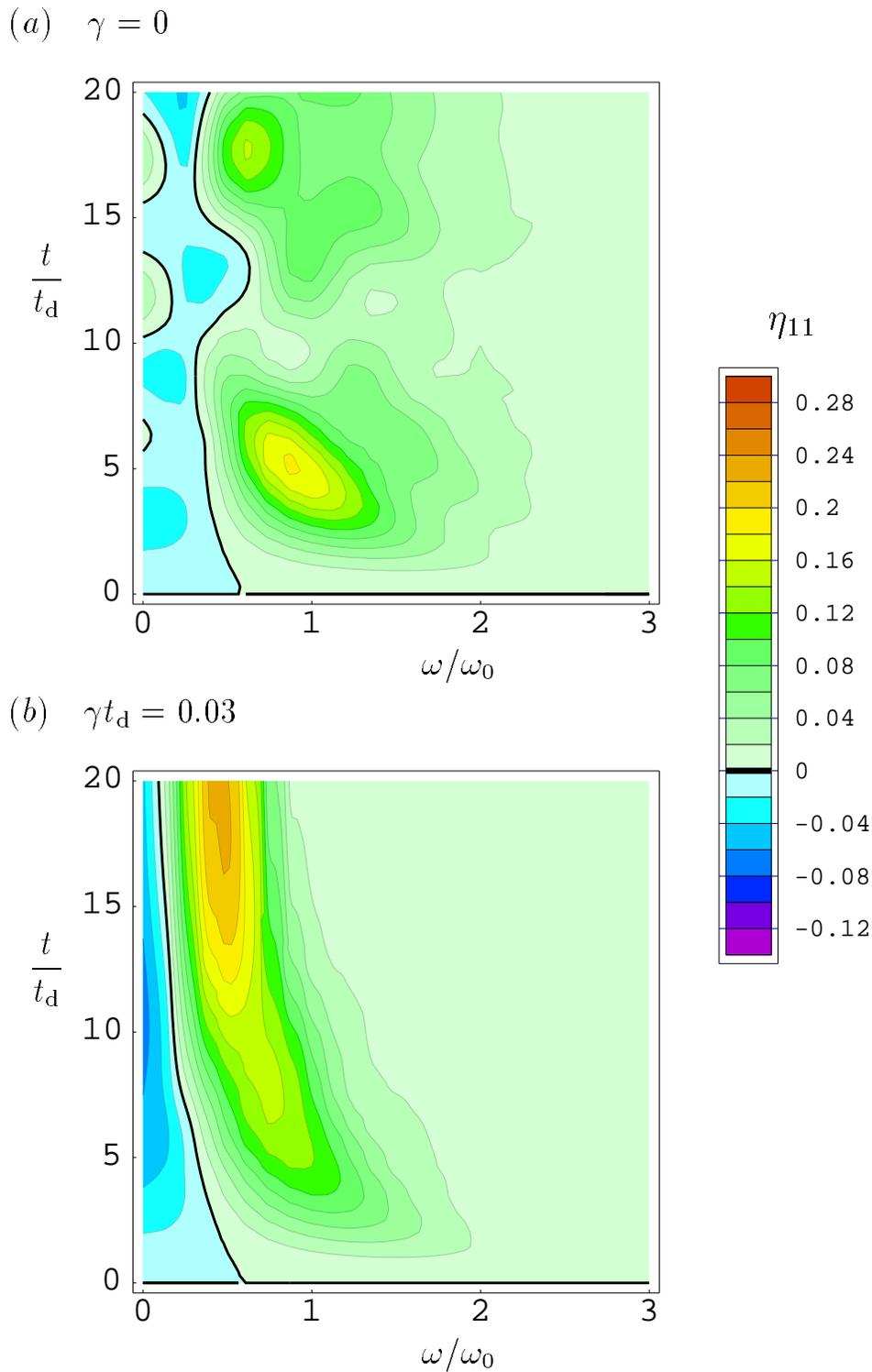,clip=}
\caption{
The temporal evolution of the coefficient $\eta _{11}$,
Eq.~(\protect\ref{eq.eta11}),
is shown as a function of the frequency $\omega$ for the
spectral interval
$(\Omega _{1},\Omega _{1}^{\prime })$ $\!=$
\mbox{$\!(\omega$ $\!-$ $\!\Delta \omega,$
$\!\omega$ $\!+$ $\!\Delta\omega$)},
$\Delta\omega$ $\!=$ $\!0.05\pi\omega_0$, and
the damping parameter $\gamma$ $\!=$ $\!0.03 /t_{\rm d}$.
Note that $\eta _{11}$ is an even function of $\omega$.
The thick solid line from the bottom to the top defines
the frequency $\Omega_0$ [used in Fig.~\protect\ref{fig.eta11x3}$(b)$]
such that $\eta_{11}$ $\!>$ $\!0$ for $|\omega|$ $\!>$ $\!\Omega_0$.
}
\label{fig.etaxt}
\end{figure}

In Fig.~\ref{fig.nn0t} the temporal evolution of the coefficient
$\eta _{11}$, Eq.(\ref{eq.eta11}), is plotted for
a narrow bandwidth mid-interval
$(\Omega _{1},\Omega _{1}^{\prime })$ $\!=$
$(-\Delta \omega,\Delta \omega)$,
$\Delta\omega/\omega_0$ $\!\ll$ $\!1$,
and various values of the damping parameter $\gamma$
(here and in the following frequencies are scaled by
$\omega_0$ $\!=$ $x_0^{-1}$ \cite{FreqNormalization}).
The strongest sub-Poissonian effect
is observed for $\gamma t_{{\rm d}}$ $\!=$ $\!0.03$ at
$t$ $\!\sim$ $\!10t_{{\rm d}}$. Surprisingly,
the sub-Poissonian statistics of the narrow bandwidth
mid-component can be stronger for a damped
soliton than for the undamped one \cite{SchmidtE98}.
The temporal evolution of the coefficient $\eta _{11}$
as a function of the frequency $\omega$ [for a narrow bandwidth
interval $(\Omega _{1},\Omega _{1}^{\prime })$ $\!=$
\mbox{$\!(\omega$ $\!-$ $\!\Delta \omega,$
$\!\omega$ $\!+$ $\!\Delta\omega$)}] is plotted in
Fig.~\ref{fig.etaxt} for
$(a)$ $\gamma$ $\!=$ $\!0$ and $(b)$ $\gamma t_{{\rm d}}$ $\!=$ $\!0.03$.
   From the figure it is seen that the interference
pattern which is typically obtained in the limit of vanishing
absorption [Fig.~\ref{fig.etaxt}$(a)$] is not observed for an
absorbing fiber [Fig.~\ref{fig.etaxt}$(b)$].
Note that in the limit of vanishing absorption the results 
obtained here are in good agreement with those obtained by 
stochastic simulations within the frame of the positive $P$ 
representation \cite{WernerMJ97b}.

Disregarding absorption and solving the cumulant evolution
equations given in \cite{SchmidtE98} in a linearization
approximation, the solution can be given by
eigenfunction expansion, the expansion coefficients
being determined by the chosen initial condition.
The result of superposition then corresponds to an interference
pattern like that in Fig.~\ref{fig.etaxt}$(a)$.
Obviously, the (phase-sensitive) terms that are superimposed do not respond
uniformly to absorption, so that the internal
coherences responsible for the interference pattern
can be destroyed at least in part.
In particular, near the center of the spectrum
\mbox{($\omega$ $\!\to$ $\!0$)} the super-Poissonian
peaks are fully suppressed and in place of them sub-Poissonian
statistics is observed. Accordingly, the super-Poissonian side-band
formation that appears with increasing frequency is more
uniform for an absorbing fiber than for a nonabsorbing one.

\begin{figure}[tbp]
\psfig{file=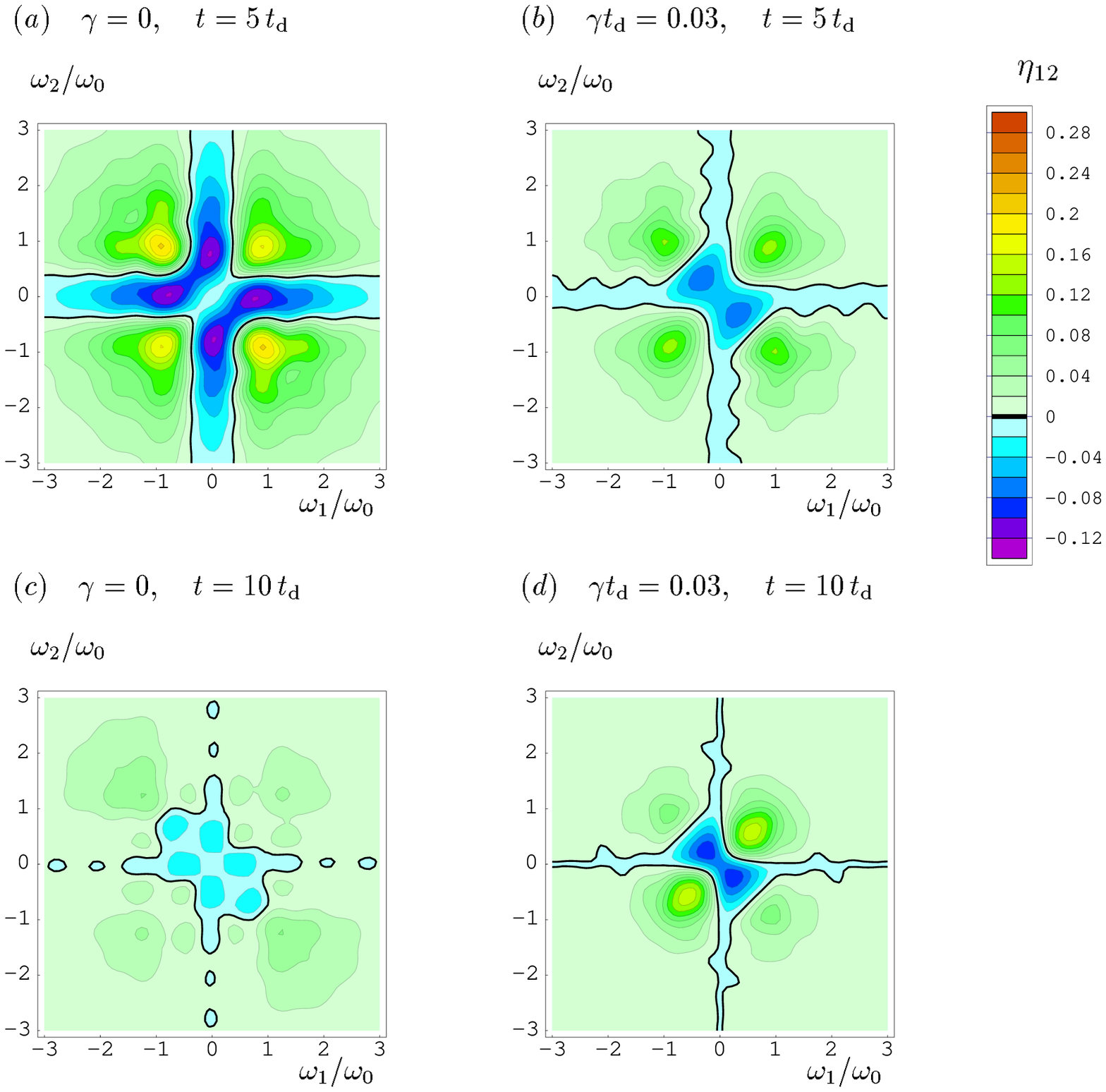,width=7in,clip=}
\caption{
The correlation coefficient $\eta _{12}$,
Eq.~(\protect\ref{eq.eta12}),
is shown as a function of the frequencies $\omega_i$
for the spectral intervals
$(\Omega _{i},\Omega _{i}^{\prime })$ $\!=$
\mbox{$\!(\omega_i$ $\!-$ $\!\Delta \omega,$
$\!\omega_i$ $\!+$ $\!\Delta\omega$)},
$\Delta\omega$ $\!=$ $\!0.05\pi\omega_0$
($i$ $\!=$ $\!1,2$), and
$(a)$ $\gamma$ $\!=$ $\!0$, $t$ $\!=$ $\!5 t_{\rm d}$,
$(b)$ $\gamma$ $\!=$ $\!0.03 /t_{\rm d}$, $t$ $\!=$ $\!5 t_{\rm d}$,
$(c)$ $\gamma$ $\!=$ $\!0$, $t$ $\!=$ $\!10 t_{\rm d}$,
$(d)$ $\gamma$ $\!=$ $\!0.03 /t_{\rm d}$, $t$ $\!=$ $\!10 t_{\rm d}$.
}
\label{fig.etaww}
\end{figure}

\begin{figure}[tbp]
\psfig{file=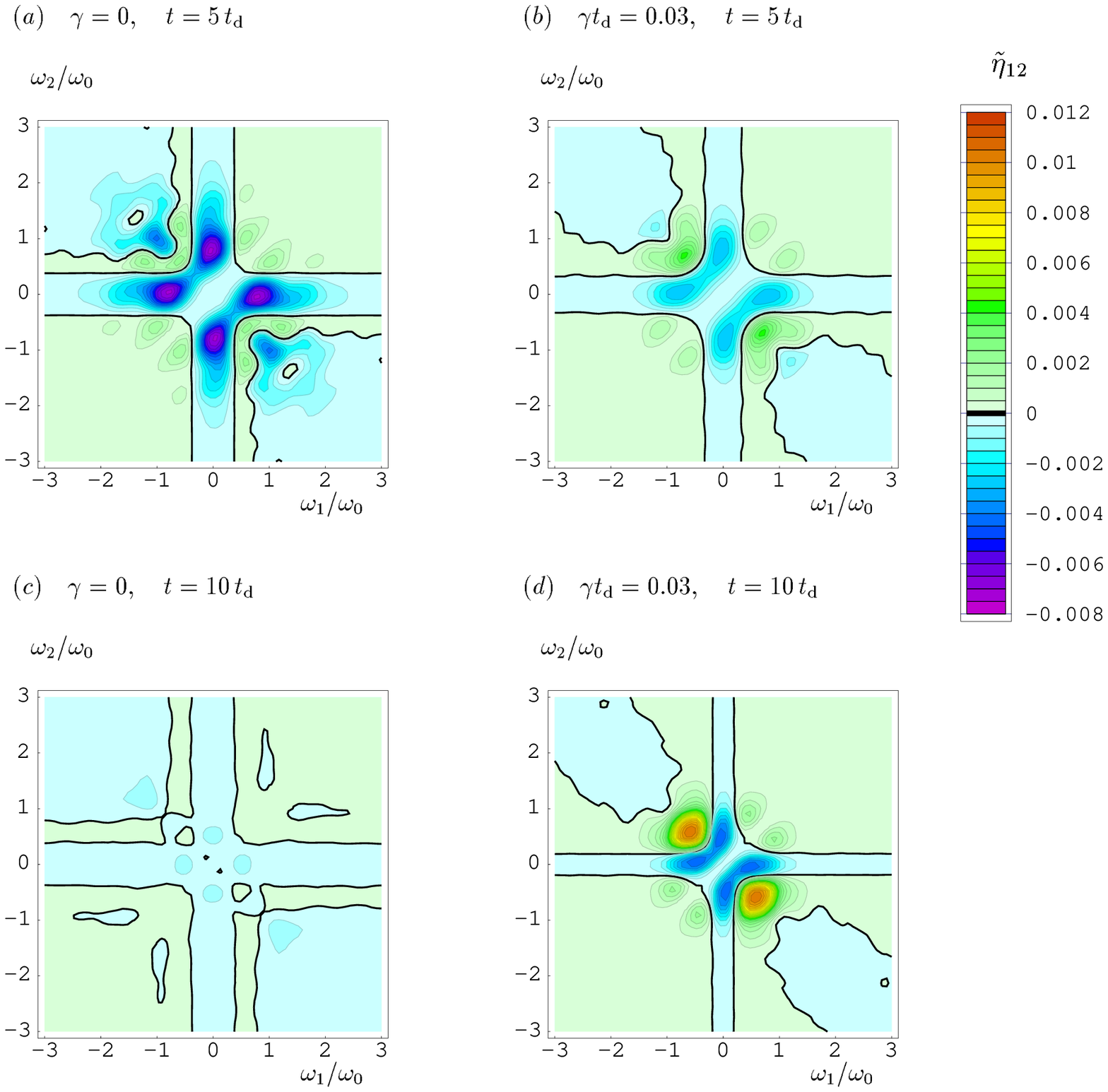,width=7in,clip=}
\caption{
The coefficient $\tilde{\eta} _{12}$,
Eq.~(\protect\ref{eq.y12}),
is shown as a function of the frequencies $\omega_i$
for the spectral intervals
$(\Omega _{i},\Omega _{i}^{\prime })$ $\!=$
\mbox{$\!(\omega_i$ $\!-$ $\!\Delta \omega,$
$\!\omega_i$ $\!+$ $\!\Delta\omega$)},
$\Delta\omega$ $\!=$ $\!0.05\pi\omega_0$
($i$ $\!=$ $\!1,2$), and
$(a)$ $\gamma$ $\!=$ $\!0$, $t$ $\!=$ $\!5 t_{\rm d}$,
$(b)$ $\gamma$ $\!=$ $\!0.03 /t_{\rm d}$, $t$ $\!=$ $\!5 t_{\rm d}$,
$(c)$ $\gamma$ $\!=$ $\!0$, $t$ $\!=$ $\!10 t_{\rm d}$,
$(d)$ $\gamma$ $\!=$ $\!0.03 /t_{\rm d}$, $t$ $\!=$ $\!10 t_{\rm d}$.
Negative values of $\tilde{\eta} _{12}$ indicate nonclassical
correlation of the photon-number variances at chosen frequencies
$\omega_1$ and $\omega_2$.
}
\label{fig.cs02dnww}
\end{figure}

\begin{figure}[tbp]
\psfig{file=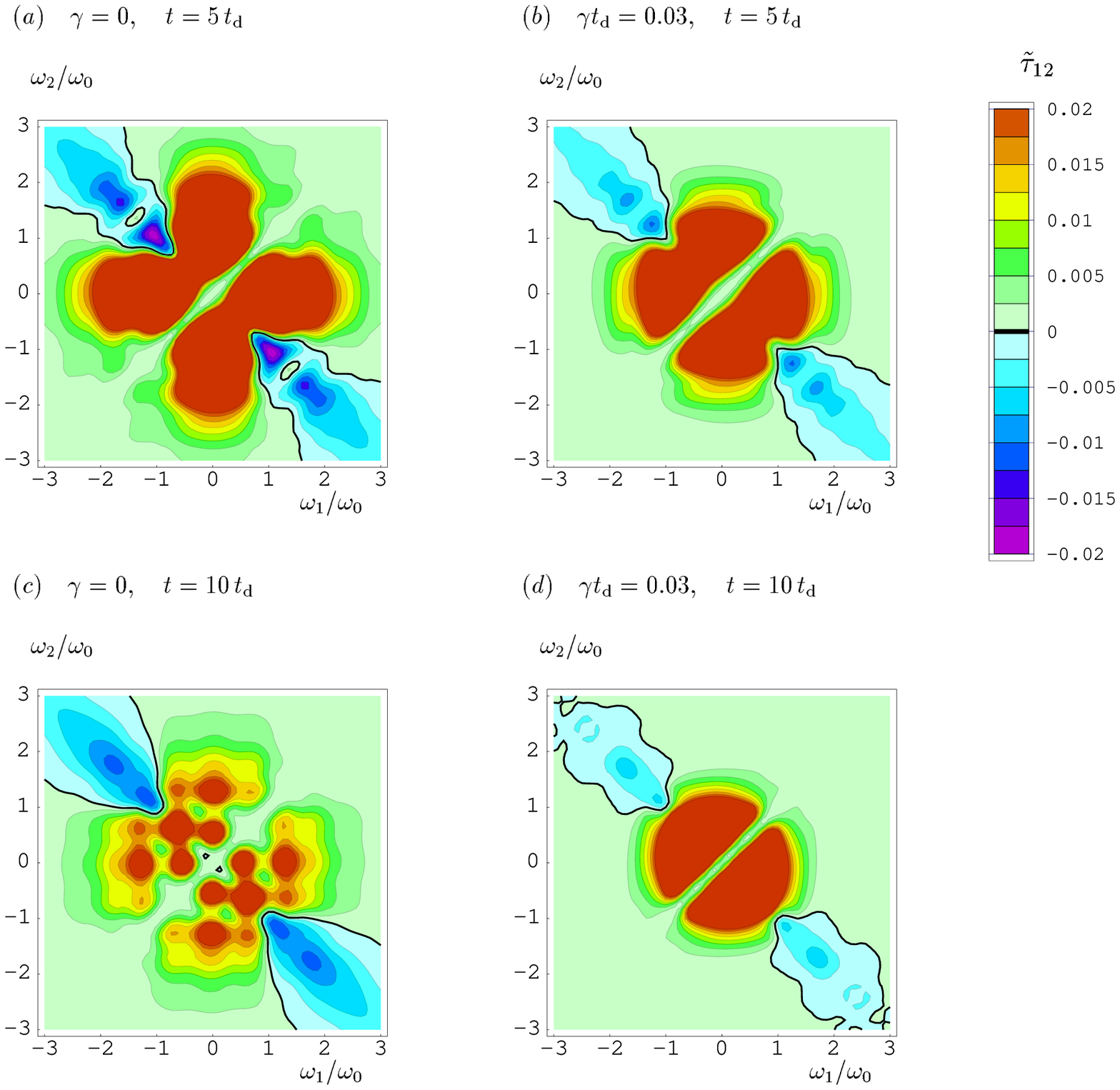,width=7in,clip=}
\caption{
The coefficient $\tilde{\tau} _{12}$,
Eq.~(\protect\ref{eq.Y12}),
is shown as a function of the frequencies $\omega_i$
for the spectral intervals
$(\Omega _{i},\Omega _{i}^{\prime })$ $\!=$
\mbox{$\!(\omega_i$ $\!-$ $\!\Delta \omega,$
$\!\omega_i$ $\!+$ $\!\Delta\omega$)},
$\Delta\omega$ $\!=$ $\!0.05\pi\omega_0$
($i$ $\!=$ $\!1,2$), and
$(a)$ $\gamma$ $\!=$ $\!0$, $t$ $\!=$ $\!5 t_{\rm d}$,
$(b)$ $\gamma$ $\!=$ $\!0.03 /t_{\rm d}$, $t$ $\!=$ $\!5 t_{\rm d}$,
$(c)$ $\gamma$ $\!=$ $\!0$, $t$ $\!=$ $\!10 t_{\rm d}$,
$(d)$ $\gamma$ $\!=$ $\!0.03 /t_{\rm d}$, $t$ $\!=$ $\!10 t_{\rm d}$.
Values $\tilde\tau_{12}$ $\!>$ $\!0.02$ are not depicted.
Negative values of $\tilde{\tau} _{12}$ indicate nonclassical
correlation of the photon numbers at chosen frequencies
$\omega_1$ and $\omega_2$.
}
\label{fig.cs02ww}
\end{figure}

To get an insight into the correlation between spectral
components at different frequencies $\omega_1$ and $\omega_2$ 
[for narrow bandwidth intervals
$(\Omega _{i},\Omega _{i}^{\prime })$ $\!=$
\mbox{$\!(\omega_i$ $\!-$ $\!\Delta \omega,$
$\!\omega_i$ $\!+$ $\!\Delta\omega$)}, $i$ $\!=1,2$],
we have plotted the correlation coefficients $\eta _{12}$ 
[Eq.~(\ref{eq.eta12}), Fig.~\ref{fig.etaww}], 
$\tilde\eta_{12}$ [Eq.~(\ref{eq.y12}), Fig.~\ref{fig.cs02dnww}], 
and $\tilde\tau_{12}$ [Eq.~(\ref{eq.Y12}),   
Fig.~\ref{fig.cs02ww}] as functions of $\omega_1$ and $\omega_2$
for \mbox{$\gamma$ $\!=$ $\!0$} and \mbox{$\gamma t_{\rm d}$ 
$\!=$ $\!0.03$}, restricting our attention
to the two characteristic propagation times \mbox{$t$ $\!=$ 
$\!5\,t_{\rm d}$} and \mbox{$t$ $\!=$ $\!10\,t_{\rm d}$}
(cf. Fig.~\ref{fig.etaxt}). 
Figures \ref{fig.etaww}$(a)$ and \ref{fig.etaww}$(b)$ reveal
that the super-Poissonian side-band components
observed at the propagation time \mbox{$t$ $\!=$ $\!5\, t_{\rm d}$}
(Fig.~\ref{fig.etaxt}) are strongly positively correlated with
respect to the photon-number variance, whereas
relatively strong negative correlation between the 
super-Poissonian side-band 
components and the sub-Poissonian mid-components is observed --
effects which can be found for both a nonabsorbing fiber,
Fig.~\ref{fig.etaww}$(a)$, and an absorbing fiber,
Fig.~\ref{fig.etaww}$(b)$.
The notably reduced correlation observed at the
propagation time \mbox{$t$ $\!=$ $\!10\, t_{\rm d}$}
for a nonabsorbing fiber, Fig.~\ref {fig.etaww}$(c)$,
may be viewed as an interference effect in a similar
sense as mentioned above. From that it can be understood
that absorption can enhance correlation --
a surprising effect which can be seen from Fig.~\ref {fig.etaww}$(d)$.

   From Fig.~\ref{fig.cs02dnww} it is seen that nonclassical 
correlation of the photon-number variance appears between 
sideband super-Poissonian 
components and -- in agreement with the condition (\ref{eq.y12.1}) --
between the sub-Poissonian mid-component and the
super-Poissonian sideband components.
The relatively strong nonclassical correlation
\mbox{$\tilde\eta_{12}$ $\!\sim$ $\!-7.3\!\cdot\!10^{-3}$}
observed for a nonabsorbing fiber at the propagation time
$t$ $\!=$ $\!5\,t_{{\rm d}}$ [Fig.~\ref{fig.cs02dnww}$(a)$]
reduces to \mbox{$\tilde\eta_{12}$ $\!\sim$ $\!-8.6\!\cdot\!10^{-4}$} at
$t$ $\!=$ $\!10\,t_{{\rm d}}$ [Fig.~\ref{fig.cs02dnww}$(c)$].
The behavior of $\tilde\eta_{12}$ for an absorbing fiber may be
quite different, as is seen from comparison of
Figs.~\ref{fig.cs02dnww}$(a)$ and \ref{fig.cs02dnww}$(c)$
with Figs.~\ref{fig.cs02dnww}$(b)$ and \ref{fig.cs02dnww}$(d)$,
respectively. Whereas at $t$ $\!=$ $\!5\,t_{{\rm d}}$
the effect of nonclassical correlation is weaker,
\mbox{$\tilde\eta_{12}$ $\!\sim$ $\! -3\!\cdot\!10^{-3}$}
[Fig.~\ref{fig.cs02dnww}$(b)$], a stronger effect is observed
at $t$ $\!=$ $\!10\,t_{{\rm d}}$,
\mbox{$\tilde\eta_{12}$ $\!\sim$ $\! -4.8\!\cdot\!10^{-3}$}
[Fig.~\ref{fig.cs02dnww}$(d)$]. The latter
reflects the enhanced correlation between the
mid-component and the sideband components as shown in
in Fig.~\ref{fig.etaww}$(d)$. We are again left with
the surprising fact that absorption can enhance
nonclassical behavior.

Figure \ref{fig.cs02ww} shows that the photon-number correlation
can be quite different from the correlation of the
photon-number variance.
Contrary to the photon-number variance,
there is no nonclassical correlation of the photon numbers
of the sub-Poissonian mid-component and the
super-Poissonian sideband components, since
for the photon number an inequality of the type
(\ref{eq.y12.1}) cannot be valid.
Nevertheless, the minimum value \mbox{$\tilde\tau_{12}$ $\!\sim$ $\! -0.02$}
obtained for a nonabsorbing fiber at the
propagation time \mbox{$t$ $\!=$ $\!5\,t_{{\rm d}}$}
[Fig.~\ref{fig.cs02ww}$(a)$]
is smaller than the minimum value
\mbox{$\tilde\eta_{12}$ $\!\sim $ $\!-7.3\!\cdot\!10^{-3}$}
[Fig.~\ref{fig.cs02dnww}$(a)$],
and hence a stronger nonclassical correlation is observed for
the photon number than the photon-number variance. 
Moreover, the effect of nonclassical correlation observed
at the propagation time \mbox{$t$ $\!=$ $\!10\,t_{{\rm d}}$} is 
weaker than that at \mbox{$t$ $\!=$ $\!5\,t_{{\rm d}}$}
for both a nonabsorbing fiber [Fig.~\ref{fig.cs02ww}$(c)$]
and an absorbing one [Fig.~\ref{fig.cs02ww}$(d)$]
[contrary to the correlation of the
photon-number variance, 
Fig.~\ref{fig.cs02dnww}$(b)$ and $(d)$]. 

\subsection{Finite bandwidth intervals}
\label{finite}

\begin{figure}[tbp]
\psfig{file=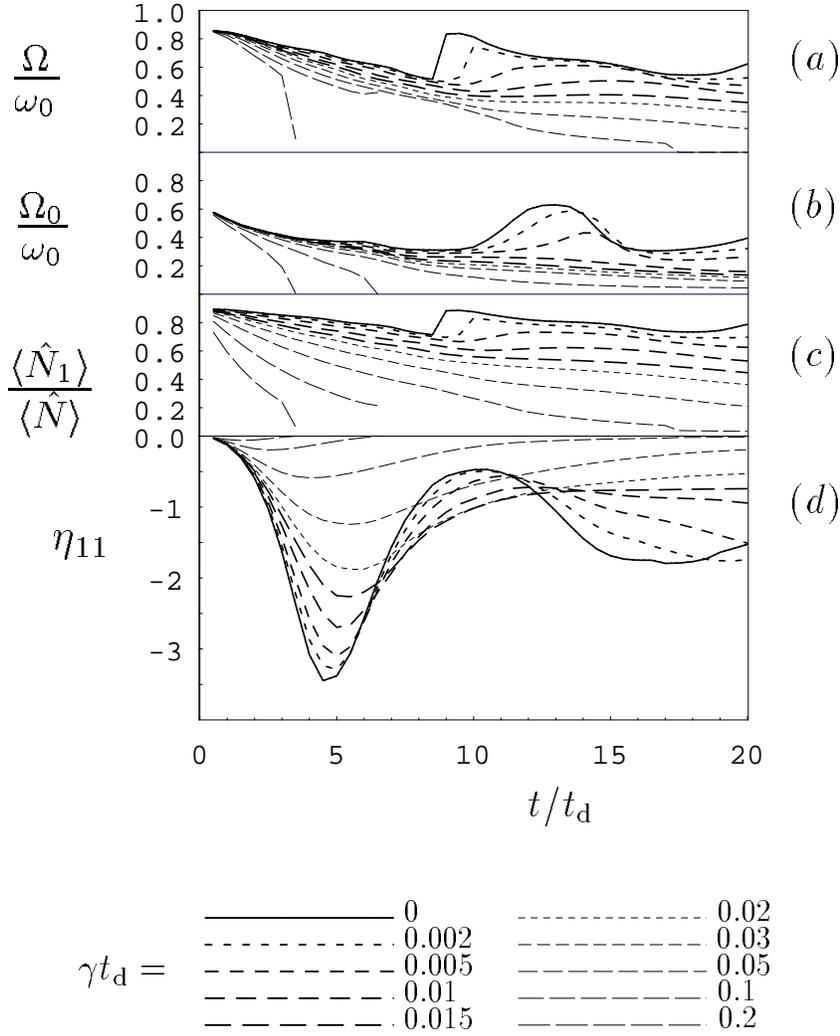,clip=}
\caption{
Optimization of the square bandpass filter
[Fig.\protect\ref{fig.setup}$(a)$] for
realization of maximum sub-Poissonian statistics.
The temporal evolution of $(a)$ the spectral interval
$(\Omega _{1},\Omega _{1}^{\prime })$ $\!=$
$\!(-\Omega,\Omega)$, $(c)$ the number $\langle\hat{N}_1\rangle$ of photons
($\langle\hat{N}\rangle$ is the total number of photons
in the initial pulse),
and $(d)$ the minimum value of the coefficient
$\eta_{11}$, Eq.~(\protect\ref{eq.eta11}),
for that interval are shown for various values of the
damping parameter $\gamma$. For comparison $(b)$, the
frequency $\Omega_{0}$ defined in Fig.~\protect\ref{fig.etaxt} is shown.
}
\label{fig.eta11x3}
\end{figure}

Let us return to the coefficient $\eta _{11}$.
Allowing for finite frequency windows, we consider
a frequency interval $(\Omega _{1},\Omega _{1}^{\prime })$ $\!=$
$\!(-\Omega,\Omega)$ [for the scheme, see Fig.~\ref{fig.setup}$(a)$]
and optimize $\Omega$ such that the sub-Poissonian components
dominates the signal and $\eta _{11}$ becomes minimal
for all times (Fig.~\ref{fig.eta11x3}). From
Fig.~\ref{fig.eta11x3}$(d)$ it is seen that
for a nonabsorbing fiber the absolute minimum of $\eta _{11}$
that is attainable in this way is $\eta _{11}$ $\!\sim$ $\!-3.5$
($t$ $\!\sim$ $\! 4.5\,t_{{\rm d}}$), which corresponds to a Fano factor
of \mbox{$F_1$ $\!=$ $\!\langle\Delta\hat{N}_1^2\rangle/
\langle\hat{N}_1\rangle$} $\!=$ $\!(1$ $\!-$ $\!\eta_{11})^{-1}$
$\!\sim$ $\!0.23$ or $\sim 6.4\,\mbox{dB}$ squeezing
(cf. \cite{WernerMJ96,MecozziA97}). In that case $\sim$ $\!80\%$ of the
photons of the initial, full pulse contribute to the
filtered pulse [Fig.~\ref{fig.eta11x3}$(c)$].
It is further seen that the oscillating behavior of
$\eta _{11}$ as a function of the propagation time is damped out
owing to absorption such that for certain
propagation times stronger sub-Poissonian statistics
can be observed for an absorbing fiber than a nonabsorbing
one. This effect is of course a consequence of the behavior
of the narrow bandwidth components as addressed in Sec.~\ref{narrow}.
The observed discrepancy between $\Omega$ [Fig.~\ref{fig.eta11x3}$(a)$]
and $\Omega _{0}$ [Fig.~\ref{fig.eta11x3}$(b)$, for the
definition of $\Omega _{0}$, see Fig.~\ref{fig.etaxt}],
e.g., for $\gamma t_{\rm d}$ $\!=$ $\!0\ldots 0.005$ at
$t$ $\!\gtrsim$ $\!9\,t_{\rm d}$, obviously results from the
correlation between different frequency components
(cf. \cite{FribergSR96,WernerMJ96}).

\begin{figure}[tbp]
\psfig{file=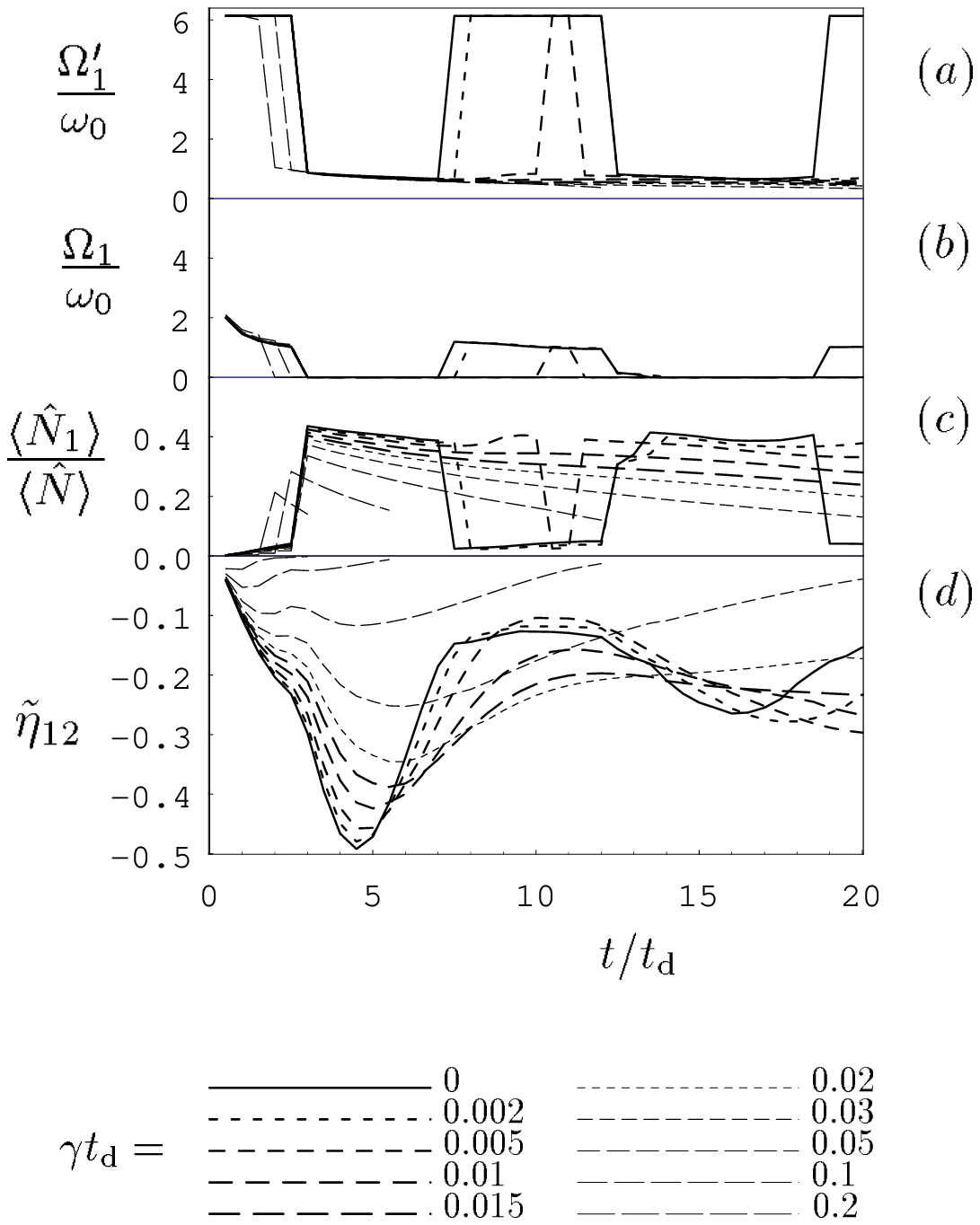,clip=}
\caption{
Optimization of two (with respect to the center of the
spectrum) symmetric square bandpass filters
[Fig.~\protect\ref{fig.setup}$(b)$]
for realization of strong nonclassical correlation
of the photon-number variance between the selected beams.
The temporal evolution of $(a),(b)$ the spectral interval
$(\Omega _{1},\Omega _{1}^{\prime })$, $(c)$ the number
$\langle\hat{N}_1\rangle$ of photons
in a beam ($\langle\hat{N}\rangle$ is the total number of photons
in the initial pulse), and $(d)$ the minimum value of the coefficient
$\tilde\eta_{12}$, Eq.~(\protect\ref{eq.y12}), are shown for various
values of the damping parameter $\gamma$.
}
\label{fig.y12nw}
\end{figure}

\begin{figure}[tbp]
\psfig{file=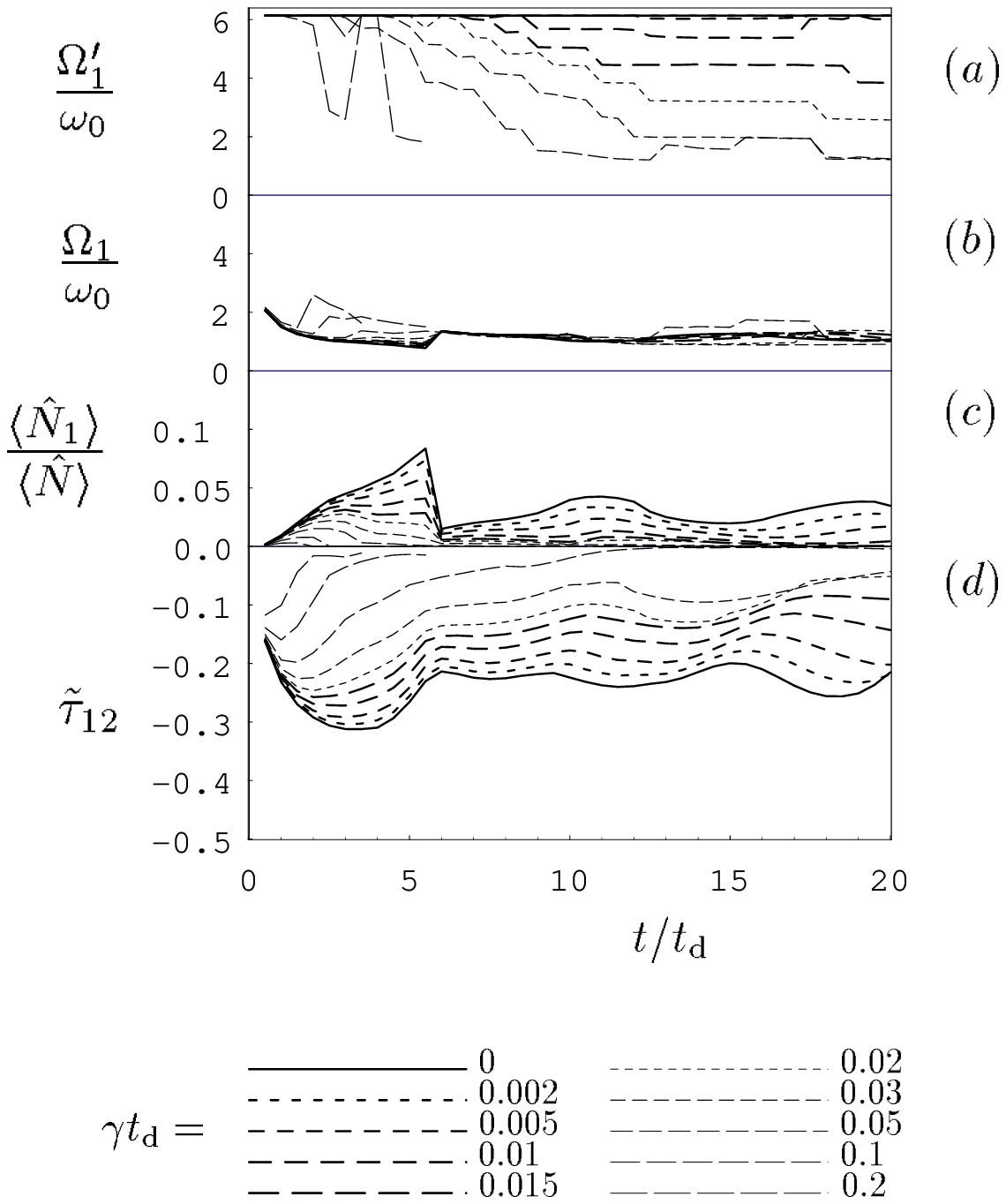,clip=}
\caption{
Optimization of two (with respect to the center of the
spectrum) symmetric square bandpass filters
[Fig.~\protect\ref{fig.setup}$(b)$]
for realization of strong nonclassical
photon-number correlation between the selected beams.
The temporal evolution of $(a),(b)$ the spectral interval
$(\Omega _{1},\Omega _{1}^{\prime })$, $(c)$ the number
$\langle\hat{N}_1\rangle$ of photons
in a beam ($\langle\hat{N}\rangle$ is the total number of photons
in the initial pulse), and $(d)$ the minimum value of the coefficient
$\tilde\tau_{12}$, Eq.~(\protect\ref{eq.Y12}), are shown for various
values of the damping parameter $\gamma$.
}
\label{fig.yy12nw}
\end{figure}

To obtain strong nonclassical correlation
between different beams, we first consider two (with respect
to the center of the spectrum)
symmetric frequency windows $(\Omega_{1},\Omega _{1}^{\prime })$ and
$(\Omega_{2},\Omega'_{2})$ $\!=$ $\!(-\Omega'_{1},-\Omega _{1})$
[for the scheme, see Fig.~\ref{fig.setup}$(b)$]
and optimize $\Omega_{1}$ and $\Omega_{1}'$
such that $\tilde\eta_{12}$ (Fig.~\ref{fig.y12nw}) and
$\tilde\tau_{12}$ (Fig.~\ref{fig.yy12nw}) become minimal. From
inspection of Fig.~\ref{fig.y12nw}$(d)$ the absolute minimum
of $\tilde\eta_{12}$ is realized for a nonabsorbing fiber
($\tilde\eta_{12}$ $\!\sim$ $\! -0.48$ at $t$ $\!\sim$ $\! 4.5\,t_{{\rm d}}$).
The effect of absorption is again seen to damp out the
oscillations of $\tilde\eta_{12}$ (as a function of the
propagation time) such that at certain propagation times
[$\gamma t_{\rm d}$ $\!=$ $\!0.01\ldots 0.02$,
$t/t_{\rm d}$ $\!=$ $\!6\ldots 14$ in Fig.~\ref{fig.y12nw}$(d)$]
stronger nonclassical correlation can be achieved for
an absorbing fiber than a nonabsorbing one.
On the contrary, Fig.~\ref{fig.yy12nw}$(d)$ reveals that
dissipation always reduces the nonclassical
photon-number correlation, i.e., for any propagation time the
lower bound of the coefficient $\tilde\tau_{12}$ is realized in
the limit of vanishing absorption. Note that the strongest
nonclassical photon-number correlation ($\tilde\tau_{12}$ $\!=$ $\!-0.32$
at $t$ $\!\sim$ $\!3.5\,t_{{\rm d}}$) is weaker than the
strongest nonclassical correlation of the photon-number variance.

For weak absorption ($\gamma t_{\rm d}$ $\!=$ $\!0\ldots 0.005$)
from Figs.~\ref{fig.y12nw}$(a)$ and $(b)$ a
(quasi-)periodic change of the frequency interval $(\Omega_{1},
\Omega _{1}^{\prime })$ is seen, in agreement with
the results in Sec.~\ref{narrow}.
At the early stage of propagation
the frequency interval is essentially determined by the
nonclassically correlated super-Poissonian sideband components.
Later [at $t$ $\!\sim$ $\!3 \,t_{{\rm d}}$ in Figs.~\ref{fig.y12nw}$(a)$
and $(b)$] it shifts towards the mid-frequency, which indicates
the increasing weight of the sub-Poissonian mid-components.
The shift back to the sideband components obviously results from the
\mbox{(quasi-)}periodic formation of super-Poissonian
(and sub-Poissonian) components in the center of the spectrum
[cf. Fig.~\ref{fig.etaxt}$(a)$].
The fraction of the number of photons in each beam relative to the
number of photons in the initial, full beam changes from less then
$5\%$ for sideband frequency windows to about of $40\%$ for
near mid-frequency windows [Fig.~\ref{fig.y12nw}$(c)$].
With increasing absorption only the first shift of the frequency window
from the sideband to the central part is observed.
In contrast to $\tilde\eta_{12}$, minimization of $\tilde\tau_{12}$
always requires sideband frequency windows [Figs.~\ref{fig.yy12nw}$(a)$
and $(b)$], where each beam contains less than $10\%$ of the initial
number of pulse photons [Fig.~\ref{fig.yy12nw}$(c)$].

\begin{figure}[tbp]
\psfig{file=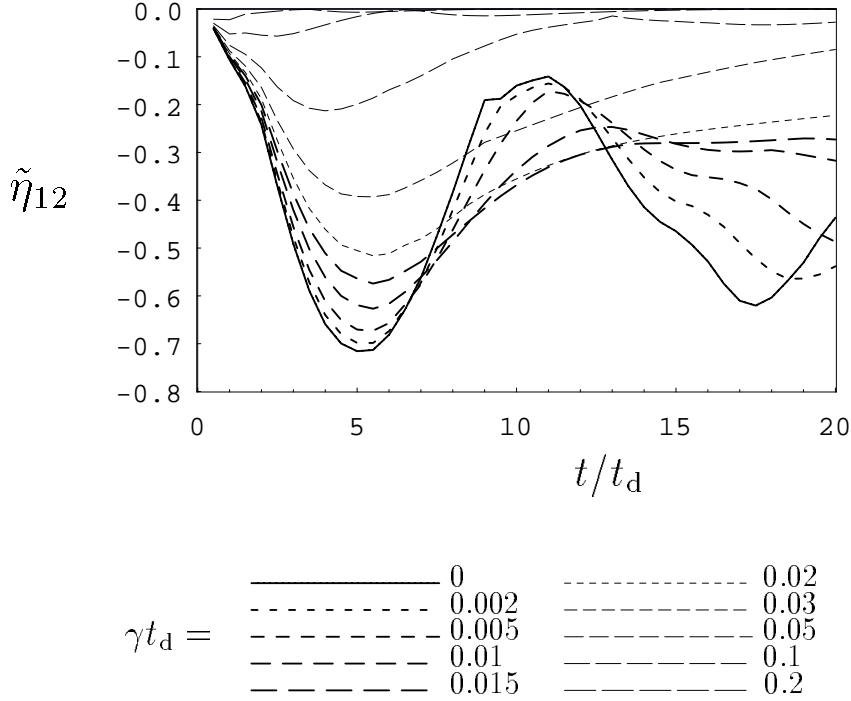,clip=}
\caption{
The temporal evolution of the minimum value of the coefficient
$\tilde\eta_{12}$, Eq.~(\protect\ref{eq.y12}), attainable for
two beams obtained from asymmetric (nonoverlapping) frequency windows
$(\Omega _{i},\Omega _{i}^{\prime })$, $i$ $\!=$ $\!1,2$,
is show for various values of the damping parameter $\gamma$.
}
\label{fig.et12}
\end{figure}

\begin{figure}[tbp]
\psfig{file=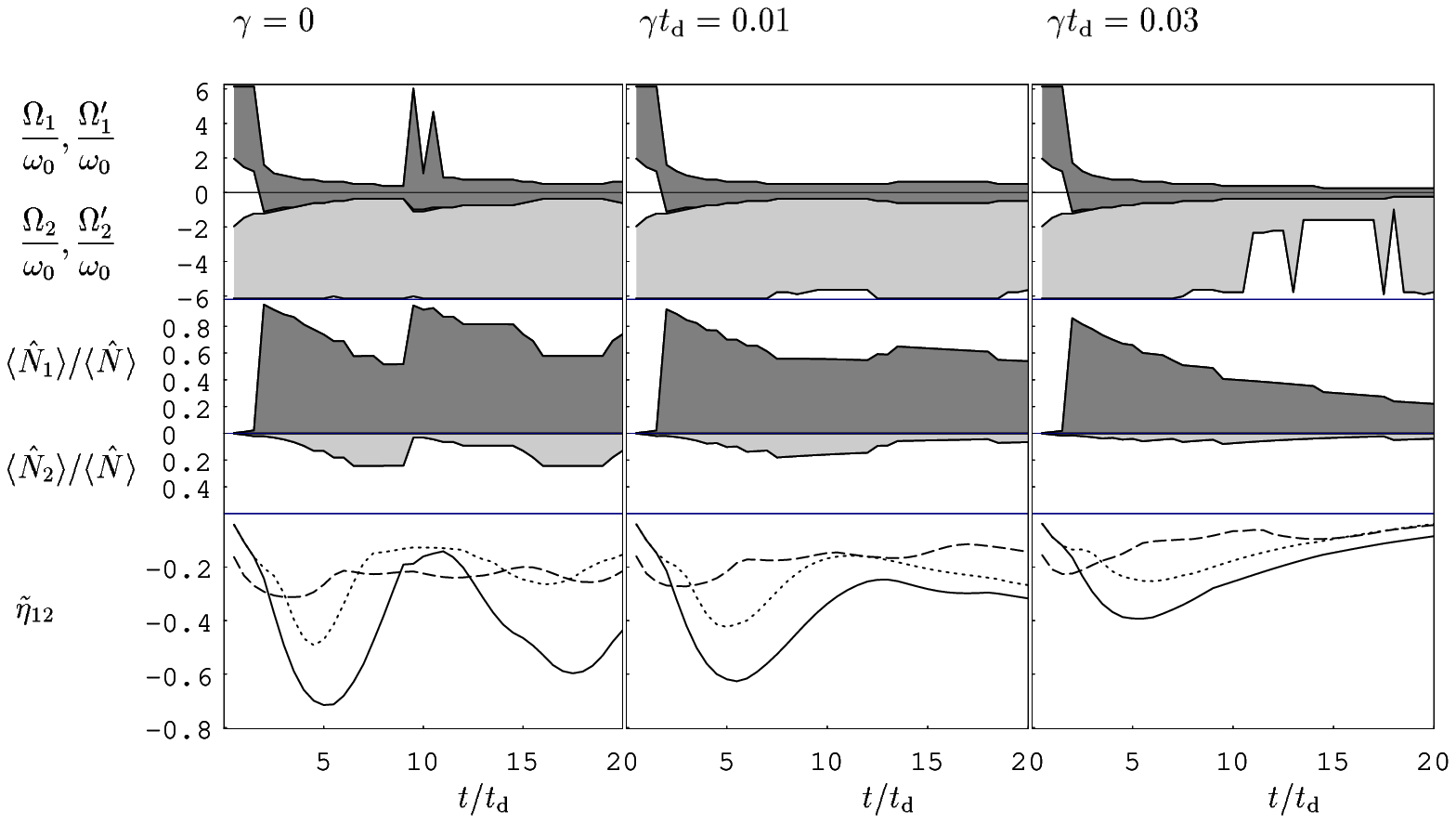,width=16cm,clip=}
\caption{
Optimization of two (with respect to the center of the
spectrum) asymmetric square bandpass filters
[Fig.~\protect\ref{fig.setup}$(b)$]
for realization of strong nonclassical correlation
of the photon-number variance between the selected beams.
The temporal evolution of the spectral intervals
$(\Omega _{i},\Omega _{i}^{\prime })$, $i$ $\!=$ $\!1,2$, the numbers
$\langle\hat{N}_i\rangle$ of photons in the beams
($\langle\hat{N}\rangle$ is the total number of photons
in the initial pulse), and the minimum value of the coefficient
$\tilde\eta_{12}$, Eq.~(\protect\ref{eq.y12}), are shown for
$(a)$ $\gamma$ $\!=$ $\!0$,
$(b)$ $\gamma$ $\!=$ $\!0.01 /t_{\rm d}$,
$(c)$ $\gamma$ $\!=$ $\!0.03 /t_{\rm d}$.
For comparison, the dotted line shows the minimum value of
$\tilde\eta_{12}$ obtained for symmetric windows as given in
Fig.~\protect\ref{fig.y12nw}, and the dashed line shows
the minimum value of $\tilde\tau_{12}$ as given in
Fig.~\protect\ref{fig.yy12nw}. Note that the symmetric
windows in Fig.~\protect\ref{fig.yy12nw} are best suited
to minimize $\tilde\tau_{12}$.
}
\label{fig.w4_15x3}
\end{figure}

Whereas symmetric windows are expected to be best suited to realization
of minimal $\tilde\tau_{12}$, from Fig.~\ref{fig.cs02dnww} it is
suggested that asymmetric windows may be more suited to realization
of minimal $\tilde\eta_{12}$. This is fully confirmed
by the calculation. The symmetric windows shown in
Fig.~\protect\ref{fig.yy12nw} indeed yield the smallest
value of $\tilde\tau_{12}$. A comparison of
Fig.~\ref{fig.y12nw}$(d)$ with Fig.~\ref{fig.et12}
reveals that $\tilde\eta_{12}$ can be reduced if asymmetric windows
are used. In particular, it is
seen that the absolute minimum of $\tilde\eta_{12}$ observed
for a nonabsorbing fiber can be reduced to
$\tilde\eta_{12}$ $\!\sim$ $\! -0.71$ ($t$ $\!\sim$ $\!5.5t_{{\rm d}}$).
A more detailed analysis is given in Fig.~\ref{fig.w4_15x3}.
It is seen that only at a early stage of propagation
symmetric and
asymmetric frequency windows yield
equal values of $\tilde\eta_{12}$. Obviously, asymmetric
frequency windows take better account of the nonclassical correlation
between sub-Poissonian mid-components and super-Poissonian
sideband components in order to reduce $\tilde\eta_{12}$.
The price to be paid are the unequal photon numbers
of the two beams, since the number of photons in the spectral
interval around the center is substantially larger
than that in the sideband interval, as is seen from the figure.
Hence for generation of (with respect to the photon-number
variance) nonclassically correlated beams, the
frequency windows should be chosen such that an
optimum between nonclassical correlation and
beam intensities is observed.

\section{Summary and concluding remarks}
\label{Sec.Summary}

We have studied the internal quantum statistics
of fundamental solitons in absorbing Kerr media, applying
multimode cumulant-expansion techniques and solving the
resulting evolution equations numerically in Gaussian approximation.
In particular, we have calculated the temporal evolution
of the photon-number variance, its correlation, and the
photon-number correlation for various frequency windows.
The formation of super-Poissonian sideband components with
nonclassically correlated photon numbers and nonclassically correlated
photon-number variances may be regarded as being a typical signature of
the quantum nature of a soliton pulse.

Since a soliton pulse is a highly involved multimode field,
interference effects can essentially determine its
properties. It is worth noting that absorption can damp out
interferences that are destructive with respect to nonclassical
features, such as squeezing and the nonclassical correlations
mentioned, so that absorption
surprisingly improves, under certain conditions,
these nonclassical effects. The calculations show that
for a nonabsorbing fiber propagation distances of
$3.5\ldots 5.5$ dispersion lengths are best suited for
detecting the nonclassical features. Destructive interferences
are observed at distances of $8\dots 12$ dispersion lengths.
At these distances the best values for photon-number squeezing
and nonclassical correlation of the photon-number variance
are achieved for an absorbing fiber ($\gamma t_{{\rm d}}$
$\!\sim$ $\!0.02$).

Using appropriately chosen spectral windows, a soliton pulse can
serve as a source of nonclassically correlated light beams.
We have considered both symmetric and asymmetric windows
and optimized them such that the filtered beams are
maximally nonclassically correlated with regard to the
photon number and the photon-number variance.
It should be pointed out that in practice a number
of effects such as Raman scattering and third-order
dispersion should be included in a refined theoretical
model. Finally, inclusion in the calculation of
non-Gaussian effects for weak absorption has been a challenge.

\acknowledgments
This work was supported by the Deutsche Forschungsgemeinschaft.
We are grateful to
F.~K\"onig,
G.~Leuchs,
and A.~Sizmann
for valuable discussions.

\widetextAAA


\AppendixBEG

\narrowtextAAA

\widetextAAA


\AppendixEND

\vspace*{2cm}

\narrowtextAAA

\vspace*{-2.5cm}


\begin{thebibliography}{10}

\bibitem{DrummondPD93}
P.~D. Drummond, R.~M. Shelby, S.~R. Friberg, and Y. Yamamoto, Nature (London)
  {\bf 365},  307  (1993).

\bibitem{SizmannA97}
A. Sizmann, Appl. Phys. B {\bf 65},  745  (1997).

\bibitem{AndersonME97}
M.~E. Anderson, D.~F. McAlister, M.~G. Raymer, and M.~C. Gupta, J. Opt. Soc.
  Am. B {\bf 14},  3180  (1997).

\bibitem{AbramI99}
I. Abram, Physics World {\bf 12},  21  (1999).

\bibitem{HasegawaA89}
A. Hasegawa, {\em Optical Solitons in Fibers} (Springer-Verlag, Berlin, 1989).

\bibitem{AkhmanovSA92}
S.~A. Akhmanov, V.~A. Vysloukh, and A.~S. Chirkin, {\em Optics of Femtosecond
  Laser Pulses} (AIP, New York, 1992).

\bibitem{KitagawaM86}
M. Kitagawa and Y. Yamamoto, Phys. Rev. A {\bf 34},  3974  (1986).

\bibitem{FribergSR92}
S.~R. Friberg, S. Machida, and Y. Yamamoto, Phys. Rev. Lett. {\bf 69},  3165
  (1992).

\bibitem{SpaelterS97a}
S. Sp{\"a}lter, P. van Loock, A. Sizmann, and G. Leuchs, Appl. Phys. B {\bf
  64},  213  (1997).

\bibitem{CourtyJM98}
J.-M. Courty {\it et~al.}, Phys. Rev. A {\bf 58},  1501  (1998).

\bibitem{RosenbluhM91}
M. Rosenbluh and R.~M. Shelby, Phys. Rev. Lett. {\bf 66},  153  (1991).

\bibitem{BergmanK94}
K. Bergman, H.~A. Haus, E.~I. Ippen, and M. Shirasaki, Opt. Lett. {\bf 19},
  290  (1994).

\bibitem{FribergSR96}
S.~R. Friberg {\it et~al.}, Phys. Rev. Lett. {\bf 77},  3775  (1996).

\bibitem{SpaelterS97}
S. Sp{\"a}lter {\it et~al.}, Europhys. Lett. {\bf 38},  335  (1997).

\bibitem{SchmittS98}
S. Schmitt {\it et~al.}, Phys. Rev. Lett. {\bf 81},  2446  (1998).

\bibitem{SpaelterS98}
S. Sp{\"a}lter {\it et~al.}, Phys. Rev. Lett. {\bf 81},  786  (1998).

\bibitem{LaiY89a}
Y. Lai and H.~A. Haus, Phys. Rev. A {\bf 40},  854  (1989).

\bibitem{YaoD95a}
D. Yao, Phys. Rev. A {\bf 52},  1574  (1995).

\bibitem{KartnerFX96}
F.~X. K{\"a}rtner and L. Boivin, Phys. Rev. A {\bf 53},  454  (1996).

\bibitem{LaiY89}
Y. Lai and H.~A. Haus, Phys. Rev. A {\bf 40},  844  (1989).

\bibitem{SingerF92}
F. Singer, M.~J. Potasek, J.~M. Fang, and M.~C. Teich, Phys. Rev. A {\bf 46},
  4192  (1992).

\bibitem{HausHA90}
H.~A. Haus and Y. Lai, J. Opt. Soc. Am. B {\bf 7},  386  (1990).

\bibitem{LaiY90}
Y. Lai and H.~A. Haus, Phys. Rev. A {\bf 42},  2925  (1990).

\bibitem{LaiY93}
Y. Lai, J. Opt. Soc. Am. B {\bf 10},  475  (1993).

\bibitem{DoerrCR94}
C.~R. Doerr, M. Shirasaki, and F.~I. Khatri, J. Opt. Soc. Am. B {\bf 11},  143
  (1994).

\bibitem{LaiY95}
Y. Lai and S.-C. Yu, Phys. Rev. A {\bf 51},  817  (1995).

\bibitem{DrummondPD87}
P.~D. Drummond and S.~J. Carter, J. Opt. Soc. Am. B {\bf 4},  1565  (1987).

\bibitem{DrummondPD93a}
P.~D. Drummond and A.~D. Hardman, Europhys. Lett. {\bf 21},  279  (1993).

\bibitem{CarterSJ95}
S.~J. Carter, Phys. Rev. A {\bf 51},  3274  (1995).

\bibitem{WernerMJ96}
M.~J. Werner, Phys. Rev. A {\bf 54},  R2567  (1996).

\bibitem{WernerMJ97}
M.~J. Werner and P.~D. Drummond, Phys. Rev. A {\bf 56},  1508  (1997).

\bibitem{WernerMJ97a}
M.~J. Werner and P.~D. Drummond, J. Comput. Phys. {\bf 132},  312  (1997).

\bibitem{WernerMJ97b}
M.~J. Werner and S.~R. Friberg, Phys. Rev. Lett. {\bf 79},  4143  (1997).

\bibitem{SchmidtE98}
E. Schmidt, L. Kn{\"o}ll, and D.-G. Welsch,
Phys. Rev. A {\bf 59}, 2442 (1999).

\bibitem{FreqNormalization}
The frequency scaling in \cite{SchmidtE98} should read
$\omega_0$ $\!=$ $\!2/x_0$, so that
$\omega_{\rm max}$ $\!=$ $\!\pi/\Delta x$ $\!=$ 
$\!5\pi\omega_0$ and $\Delta\omega_{\rm min}$ $\!=$ 
$\!2\pi/(m\Delta x)$ $\!=$ $\!0.05\pi\omega_0$
for the chosen values of $\Delta x$ $\!=$ $0.1x_0$ and 
$m$ $\!=$ $200$.

\bibitem{MecozziA97}
A. Mecozzi and P. Kumar, Opt. Commun. {\bf 22},  1232  (1997).

\end{thebibliography}


\widetextAAA

\newpage

\end{document}